\documentclass[apl,twocolumn,superscriptaddress,groupedaddress,showpacs]{revtex4}
\usepackage{graphicx}
\usepackage{amsmath}
\usepackage{amssymb}
\usepackage{wasysym}
\usepackage{pifont}

\def\Jh{\{{\cal J},h\}}
\def\frac12{{1\over 2}}

\newcommand{\E}{\mathbb{E}}

\newcommand{\R}{\mathbb{R}}

\newcommand{\N}{\mathbb{N}}

\newcommand{\beq}{\begin{equation}}
\newcommand{\eeq}{\end{equation}}

\begin{document}

\title{Thermodynamic Identities and Symmetry Breaking in Short-Range Spin Glasses}

\author{L.-P.~Arguin}
\affiliation{Department of Mathematics, City University of New York, Baruch College and Graduate Center, New York, NY 10010}
\author{C.M.~Newman}
\affiliation{Courant Institute of Mathematical Sciences, New York, NY 10012 USA and NYU-ECNU Institute of Mathematical Sciences at
NYU Shanghai, 3663 Zhongshan Road North, Shanghai 200062, China}
\author{D.L.~Stein}
\affiliation{Department of Physics and Courant Institute of Mathematical Sciences,
New York University, New York, NY 10012 USA and NYU-ECNU Institutes of Physics and Mathematical Sciences at NYU Shanghai, 3663 Zhongshan Road North, Shanghai, 200062, China}

\begin{abstract}
  We present a technique to generate relations connecting pure state
  weights, overlaps, and correlation functions in short-range spin
  glasses.  These are obtained directly from the unperturbed
  Hamiltonian and hold for general coupling distributions. All are satisfied in phases with simple thermodynamic
  structure, such as the droplet-scaling and chaotic pairs
  pictures. If instead nontrivial mixed-state pictures hold, the relations suggest that replica symmetry is broken as described by a Derrida-Ruelle
  cascade, with pure state weights distributed as a Poisson-Dirichlet process.
 \end{abstract}


\maketitle

{\it Introduction.} The thermodynamic behavior in finite dimensions of
short-range Ising spin glasses remains an open
problem~\cite{SN13}. Several pictures of the thermodynamics of the
low-temperature phase have been proposed, but analytical results are
difficult to obtain. In this paper we describe a method for generating
infinite sets of identities in short-range spin glasses for general
couplings, at any temperature and in any dimension. We will see that
they are especially useful for studying nontrivial mixed-state
pictures (including replica symmetry breaking), and we use them to
provide strong analytical evidence that, if a nontrivial mixed-state
picture exists in some dimension, then its
symmetry breaking is described by a Derrida-Ruelle cascade~\cite{Derrida85,Ruelle87,BS98}.
Full replica symmetry breaking corresponds to a $k$-step Derrida-Ruelle cascade
in the limit $k\to\infty$~\cite{Ruelle87,PRY15}.
Moreover, the pure state weights are distributed as a Poisson-Dirichlet~(PD) point process. (For
a nice discussion of the PD distribution and its relation to various
spin glass models, see~\cite{CG13}.) Such a distribution of pure state
weights has been proved for the generalized random-energy
model~(GREM)~\cite{BK07} and for certain cases of $p$-spin mean-field
spin glasses~\cite{Talagrand03} in restricted ranges of
temperature. It is believed to hold as well for the canonical
Sherrington-Kirkpatrick model~\cite{SK75}, though no proof yet
exists~\cite{Bolthausen07}. We will present the formal definition of
the PD~distribution below, but note informally here that it
corresponds to the characterization of the free energies of mean-field
spin glass states as independent random variables with an exponential
distribution~\cite{MPV85,DT85}.

{\it Proposed scenarios for the spin glass phase.} There is good experimental~\cite{BY86,GSNLAI91} and numerical~\cite{BY86,PC99,Ballesteros00,HPV08,Janus13} evidence for a phase transition in three dimensions and above. Assuming, as most do (backed by numerical studies; see for example~\cite{Ballesteros00}) that this is accompanied by broken spin-flip symmetry --- equivalently,  a nonzero Edwards-Anderson order parameter $q_{EA}$~\cite{EA75} --- there are several possibilities for the thermodynamic structure of the spin glass phase. A natural framework for discussing this is the metastate~\cite{SN13,AW90,NS96b,NSBerlin,NS97,NS98,NS02,NS03jpc,NS05a},
which is an ensemble of thermodynamic states, of which there can in principle be one or many~\cite{notemany}. Each of these thermodynamic states
comprises either a single pure state or else a convex mixture of distinct pure states; if the latter, we refer to it as a {\it mixed\/} state.
If spin-flip symmetry is present in the Hamiltonian and broken at some fixed inverse temperature $\beta$, and if the metastate is generated through a sequence of finite volumes with spin-symmetric boundary conditions (such as periodic, antiperiodic, or free), then each of the thermodynamic states in the metastate must be mixed, giving equal weight to spin-reversed pairs of pure states. If each such thermodynamic state consists of only a single pair of spin-reversed pure states, each with weight 1/2, we will refer to it as a {\it trivial mixture\/}; a nontrivial mixture denotes a thermodynamic state comprising an infinite set of pairs of pure states, with the members of each pair having equal weight and the weights of all pure states summing to unity~\cite{note1}.

We now list the main proposed possibilities for the thermodynamics of short-range spin glasses, under the assumption of broken spin-flip symmetry. The simplest picture, from the viewpoint of structure and organization of pure states, is one that conjectures a single thermodynamic state consisting of a trivial mixture of pure states. This well-known picture arises from the {\it droplet-scaling\/} ansatz, and was put forward by McMillan~\cite{Mac84}, Bray and Moore~\cite{BM}, and Fisher and Huse~\cite{FH86,FH87,FH88}. This picture asserts that the low-temperature spin glass phase is supported on a single pure state pair.

In order of increasing complexity (again from the viewpoint of pure state structure and organization), the next picture supposes an infinite ensemble of thermodynamic states, but with each such state a trivial mixture.
This is the {\it chaotic pairs\/} picture~\cite{SN13,NS03jpc,NS96b,NSBerlin,NS97,NS98,NS02}, and is a ``many-state'' picture (that is, the metastate is supported on an uncountable infinity of pure state pairs), but one in which the spin overlap structure is trivial, as in droplet-scaling.

A far more complex picture supposes an infinite ensemble of thermodynamic states, each of which is a {\it nontrivial\/} mixture of pure states. There are many possible scenarios of pure state structure and organization that are {\it a priori\/} consistent with nontrivial mixed-state pictures, but the most well-known is the {\it replica symmetry breaking\/}~(RSB) picture due to Parisi and co-workers~\cite{P79,MPSTV84,MPV87}. This picture has a complicated overlap structure.

As noted elsewhere~\cite{NS01}, another important scenario~\cite{KM00,PY00}, known as `TNT' (trivial edge,
nontrivial spin overlap), can be consistent with any of the above three (but see also~\cite{CGGV06}). There has been a large body of numerical work
attempting to resolve the question of which of these pictures  (if any) describes spin glass ordering; some 
of the more recent include~\cite{KY05,B14,YKM12,Middleton13,MG13,BMMMY14,WMK14}.

Both the droplet-scaling and the chaotic pairs scenarios exhibit a relatively simple thermodynamic structure. The RSB picture is much more complex;
the basics of its thermodynamic structure took a long time to be elucidated~\cite{SN13,NS03jpc,MPRRZ00,Read14,NS02}, and many
of its features remain to be understood. Previous papers by the authors have presented a combination of rigorous and heuristic arguments that cast doubt on the internal consistency of nontrivial mixed-state pictures for short-range spin glasses in finite dimensions~\cite{SN13,NS98,NS03jpc} (but see~\cite{Read14} for a critique of one of these); nevertheless, in the absence of a rigorous argument they remain viable, and any analytical results on their properties are therefore useful.

{\it Thermodynamic states.\/} We consider the Edwards-Anderson~(EA)~\cite{EA75}
nearest-neighbor spin glass on a $d$-dimensional cubic lattice and
(possibly) in a small applied magnetic field. The Hamiltonian is 
\begin{equation}
\label{eq:EA}
H_{\Jh}=-\sum_{<x,y>} J_{xy} \sigma_x\sigma_y - \epsilon\sum_xh_x\sigma_x,
\end{equation}
where the couplings
$J_{xy}$ and fields $h_x$ are independent,
identically distributed continuous random variables with mean zero and variance one, the first sum is over
nearest neighbors only, and $\Jh$ denotes
a particular realization of the couplings and fields.  If a magnetic field is present, $\epsilon>0$; otherwise, 
$\epsilon=0$.  

Local properties (such as correlation functions) in a given large volume correspond to a particular thermodynamic state,
denoted by $\Gamma$~\cite{SN13,NS02,NS03jpc}.  
Each $\Gamma$ must be either a countable or continuum mixture of pure
states (or a combination of the two).  We consider here only $\Gamma$'s with a countable decomposition into pure states:
\begin{equation}
\label{eq:sum}
\Gamma=\sum_{\alpha} W_{\alpha}\rho_\alpha\ ,
\end{equation}
where $W_{\alpha}=W_\alpha(\Gamma)$ is the weight of pure state
$\rho_{\alpha}$ (or simply $\alpha$) in~$\Gamma$. The spin overlap is defined as usual by
\begin{equation}
\label{eq:qab}
q^{\alpha\beta}=\lim_{L_0\to\infty}|\Lambda_{L_0}|^{-1} \sum_{x\in\Lambda_{L_0}}
\langle\sigma_x\rangle_\alpha \langle\sigma_x\rangle_\beta\, , 
\end{equation}
where $\langle\cdot\rangle_\alpha$ denotes a thermal average in pure state
$\alpha$.  Edge overlaps are defined similarly. We assume in what follows that at fixed temperature 
the self-overlap $q^{\alpha\alpha}=q_{EA}$ of every pure state is identical~\cite{magnote}.

{\it Thermodynamic relations.\/} A useful tool for obtaining our result is the {\it metastate\/}, discussed
in~\cite{SN13,AW90,NS96b,NSBerlin,NS97,NS98,NS02,NS03jpc,NS05a}.  The metastate~$\kappa_{\Jh}$ can be thought of as a probability measure on thermodynamic
states $\Gamma$ induced by an infinite sequence of volumes with specified boundary conditions.  It contains all thermodynamic
information that can be generated in the ensemble of all volumes.

Our main result is the following: For the EA Hamiltonian~(\ref{eq:EA}), consider a metastate
$\kappa_{\Jh}$ constructed using coupling-independent boundary conditions, such as periodic.  Then an infinite set of thermodynamic identities connecting weights, overlaps, and correlation functions, valid in any finite dimension and at any temperature, can be constructed using a well-defined procedure. 
These can be used to restrict possible scenarios of a spin glass phase. In this paper we use these relations to provide strong analytical evidence for the following conclusion: if the metastate is such that its $\Gamma$'s are supported on countably infinite mixtures of distinct pure states, then the weights of those pure states are distributed according to a Poisson-Dirichlet point process.

We consider below the case $\epsilon>0$ (so pure states no longer appear in the $\Gamma$'s within spin-reversed pairs of equal weight).
As a consequence the derived relations involve spin rather
than edge overlaps. We return briefly to the case of $\epsilon=0$ later. 
We choose periodic boundary conditions, although
the argument is equally valid for other choices, such as free or fixed.

Before proceeding, we define some quantities that will be needed later. Let
$\langle\sigma_x\rangle_\alpha$ be the thermal expectation of the spin
at $x$ in pure state $\alpha$, and
$\langle\sigma_x\rangle_\Gamma=\sum_\alpha
W_\alpha\langle\sigma_x\rangle_\alpha$ be its expectation in the mixed
state $\Gamma$. We also define the average overlap $q_\Gamma$ in
$\Gamma$ (for ease of notation, dependence on $\Jh$ is hereafter
dropped):
\begin{equation}
\label{eq:qgamma1}
q_\Gamma=\sum_{\alpha\beta}W_\alpha W_\beta q^{\alpha\beta}\, .
\end{equation}
It follows that
\begin{equation}
\label{eq:qgamma2}
q_\Gamma=\lim_{L_0\to\infty}|\Lambda_{L_0}|^{-1}\sum_{x\in\Lambda_{L_0}}\langle\sigma_x\rangle_\Gamma^{\ 2}\, .
\end{equation}
It is clear from~(\ref{eq:qgamma2}) that $q_\Gamma$ is nonnegative.
It also follows from a Cauchy-Schwartz inequality that if $\Gamma$
comprises more than one pure state, then $q_\Gamma<q_{EA}$.  

Now for some $\Jh$, and at fixed inverse temperature~$\beta$ and fixed $\epsilon$, we choose an arbitrary site $x$ and change the
applied field at that site: $h_x\rightarrow h_x'=h_x +\Delta h_x$.  In any
$\Gamma$, every pure state $\alpha$ transforms~\cite{AW90} (see
also~\cite{NSBerlin,NS97,NS98,NS02,NS03jpc}) to a new pure state $\alpha'$, with
\begin{eqnarray}
\label{eq:walpha}
W_\alpha\to W_{\alpha'}=r_\alpha W_\alpha/\sum_\gamma r_\gamma W_\gamma\\
\label{eq:ralpha}
r_\alpha=\Bigl\langle\exp(\beta \epsilon \Delta h_x\sigma_x)\Bigr\rangle_\alpha\, .
\end{eqnarray}
Then
\begin{equation}
W_{\alpha'}=W_\alpha{\Bigl(1+\tanh(\beta\epsilon\Delta h_x)\langle\sigma_x\rangle_\alpha\Bigr)\over\Bigl(1+\tanh(\beta\epsilon\Delta h_x)\langle\sigma_x\rangle_\Gamma\Bigr)}\, ,
\end{equation}
where $\langle\sigma_x\rangle_\alpha$, $\langle\sigma_x\rangle_\Gamma$,
and $W_\alpha$ are all evaluated at $\Delta h_x=0$. 
Using the notation
$\partial_x\equiv\lim_{\Delta h_x\to 0}\partial/\partial(\beta\epsilon\Delta h_x)$ and $\partial_{xx}\equiv\lim_{\Delta h_x\to 0}\partial^2/\partial(\beta\epsilon\Delta h_x)^2$, we have
\begin{equation}
\label{eq:firstderiv}
\partial_xW_\alpha = W_\alpha \Bigl(\langle\sigma_x\rangle_\alpha - \langle\sigma_x\rangle_\Gamma\Bigr)
\end{equation}
and
\begin{equation}
\label{eq:secondderiv}
\partial_{xx}W_\alpha = 2W_\alpha \Bigl(\langle\sigma_x\rangle_\Gamma^2 -\langle\sigma_x\rangle_\alpha\langle\sigma_x\rangle_\Gamma\Bigr)\, .
\end{equation}

Now consider a function $f(\Gamma)$ on the states $\Gamma$ (equivalently, on the
correlation functions that characterize $\Gamma$). Let $\E[f]$ denote the average
of $f$ over the metastate (as always, for a single, fixed realization $\Jh$); that is, 
\begin{equation}
\label{eq:fav}
\E[f]=\E_{\kappa_{\Jh}} [f] = \int{d\kappa_{\Jh}(\Gamma)}\ f(\Gamma)\, .
\end{equation}
If $f$ is a measurable, translation-invariant function of both the coupling and field realizations,
then by the spatial ergodic theorem $\E[f]$ must be the same for almost every
$\Jh$ (see, for example,~\cite{NS96a}).

To derive useful relations, one may consider many different measurable, translation-invariant $f(\Gamma)$'s, each of which
leads to different identities. Here we consider 
\begin{equation}
\label{eq:f}
f(\Gamma)={1\over n}\log\Bigl(\sum_\alpha W_\alpha^n\Bigr)\, ,
\end{equation}
where $n$ is any real number greater than one. By the above arguments, the metastate integral $\int_{\kappa_{\Jh}} f(\Gamma)$ 
is constant~a.s.~with respect to ${\Jh}$. 

%
%

We now compute the second derivative with respect to $\beta\epsilon\Delta h_x$ of the integral~(\ref{eq:fav}), using~(\ref{eq:f}) as the integrand. Taking the limit $\Delta h_x\to 0$ of the result, averaging over all sites $x$, and using
the required constancy of the metastate integral of $f$, we find~\cite{notederiv}
\begin{eqnarray}
\label{eq:intderiv}
0&=&\lim_{L_0\to\infty}|\Lambda_{L_0}|^{-1}\sum_{x\in\Lambda_{L_0}}\ {\partial_{xx} \E[f(\Gamma)}]\\
&=&\E\Big[(n-1)q_{EA} +\ q_\Gamma - n\sum_{\alpha\beta}p^{(n)}_\alpha p^{(n)}_\beta q^{\alpha\beta}\Big]\nonumber\, ,
\end{eqnarray}
where $p^{(n)}_\alpha = W_\alpha^n/\sum_\alpha W_\alpha^n$ depends on $\Gamma$.  

As noted,~(\ref{eq:intderiv}) is only one of many possible sets of thermodynamic relations that must be satisfied by any thermodynamic phase arising from the EA~Hamiltonian~(\ref{eq:EA}). All such relations
are trivially satisfied by both the droplet-scaling and chaotic pairs pictures: when $\epsilon>0$, both consist of $\Gamma$'s (a single $\Gamma$ in the case of droplet-scaling, many in the case of chaotic pairs) comprising a single pure state. 
The situation is very different for nontrivial mixed-state pictures: now the identities provide strong constraints on the relations between weights, overlaps, and/or correlation functions. 

Rewriting~(\ref{eq:intderiv}), we find 
\begin{equation}
\label{eq:simplified}
1-\E[\bar q_{\Gamma}]=n\E\Big[1-\sum_{\alpha,\beta}p^{(n)}_\alpha p_\beta^{(n)} \bar{q}^{\alpha\beta}\Big]\, ,
\end{equation}
where $\bar{q}^{\alpha\beta}=q^{\alpha\beta}/q_{EA}$ and similarly for $\bar q_\Gamma$. The RHS of~(\ref{eq:simplified}) must be independent of $n$. 
A similar relation was found for the SK model~\cite{MPV85,DT85}  and already rules out many possible distributions of the weights.


{\it Derrida-Ruelle cascades and Poisson-Dirichlet distributions.\/}  The Derrida-Ruelle cascades are the precise mathematical formulation of replica symmetry breaking: a cascade of $k$~levels corresponds 
to $k$-step~RSB. They are based on Poisson point processes with exponential density. 
Consider the set $(F_\alpha,\alpha\in \N)$, a Poisson process with density $\exp(Cx)dx$. Physically, in a volume with $L^d$ spins each $F_\alpha$ corresponds to the $O(L^{-d})$ correction to the free energy per spin of the pure state $\alpha$ within a $\Gamma$ taken from the metastate. For simplicity we have set the minimum of the $F_\alpha$'s to zero.

The set of corresponding Gibbs weights~$e^{-\beta F_\alpha}$ for the pure states at inverse temperature~$\beta$  is also Poisson with density~$({C\over \beta}) y^{-C/\beta -1}dy$ on~$\R$.
If $F_\alpha$ is Poisson with density~$\exp(Cx)dx$, then given that $W_\alpha=e^{-\beta F_\alpha}/\sum_{\gamma} e^{-\beta F_\gamma}$, the collection $W_\alpha$ is not Poisson, because of the dependence introduced by the normalization. The distribution for this collection of weights is called {\it Poisson-Dirichlet} with parameter $\lambda$, or simply $PD(\lambda)$. In the REM, $\lambda=\beta_c/\beta<1$~(see, for example, \cite{Talagrand03}), and more generally it is a function of both temperature and field. It is not hard to see that, if $W_\alpha$ is $PD(\lambda)$, then $p^{(n)}_\alpha$ is $PD(\lambda/n)$ for any $n>1$. 

We can now compute the density of the weights on $[0,1]$. It was shown in~\cite{Ruelle87,MPV85,DT85} that the density of states is
\begin{equation}
\label{eq:dos}
\E(\#\{\alpha: W_\alpha \in [u,u+du]\})= {(1-u)^{\lambda-1}u^{-\lambda-1}\over\Gamma(\lambda)\Gamma(1-\lambda)}du\, ,
\end{equation}
which implies (for example) that $\E[\sum_\alpha W_\alpha^2]=1-\lambda$.

{\it 1-step RSB.\/} This is usually understood as ${q}^{\alpha\beta}=q\ (<q_{EA})$ for all $\alpha\ne\beta$.
In this case $\bar q_\Gamma=\bar q+ (1-\bar q)\sum_\alpha W_\alpha^2$. If $W_\alpha$ is PD($\lambda$), then
$\E[\bar q_{\Gamma}]=1-\lambda(1-\bar q)$, and 
\begin{eqnarray}
\label{eq:1step}
\E\Big[\sum_{\alpha,\beta}p^{(n)}_\alpha p_\beta^{(n)} \bar{q}_{\alpha\beta}\Big]&=&\E\Big[\sum_{\alpha}(p^{(n)}_\alpha)^2\Big]+\bar q\E\Big[\sum_{\alpha\neq \beta}p^{(n)}_\alpha p_\beta^{(n)}\Big]\nonumber\\
&=&1-\lambda/n+\bar q\lambda/n\, ,
\end{eqnarray}
which satisfies the consistency relations~(\ref{eq:simplified}) for all $n$.

{\it k-step RSB.\/} The consistency of PD distributions for the weights with 2-step~RSB applied to~(\ref{eq:simplified}) is shown in detail in the Supplementary~Notes~\cite{supp}.
Here we consider the general $k$-step case. Eq.~(\ref{eq:simplified}) can be written as
\begin{equation}
\label{eq:k1}
1-\int_0^1 q dx(q)=n\left(1-\int_0^1 q dx^{(n)}(q)\right)
\end{equation}
where $x(q)$ is the cumulative distribution function (cdf) of the 2-replica spin overlap:~$x(q)=\E[\sum_{\alpha,\beta}W_\alpha W_\beta \delta(q_{\alpha\beta}-q)]$, and $x^{(n)}(q)$ is the cdf of the 2-replica spin overlap distribution after applying the map sending $W_\alpha$ to $p_\alpha^{(n)}$. We can integrate~(\ref{eq:k1}) by parts to get 
\begin{equation}
\label{eq:k2}
\int_0^1x(q) \ dq=n\int_0^1x^{(n)}(q) \ dq\ .
\end{equation}
For $k$-step RSB, the states can be labeled by $\alpha=(\alpha_1,\dots,\alpha_k)$, with weights given by
\begin{equation}
\label{eq:k3}
W_\alpha=e^{-\beta F_{\alpha_1}}\dots e^{-\beta F_{\alpha_1\dots \alpha_k}}/\sum_{\gamma_1,\dots,\gamma_k}e^{-\beta F_{\gamma_1}}\dots e^{-\beta F_{\gamma_1\dots \gamma_k}}\, ,
\end{equation}
where $(F_{\alpha_1\dots \alpha_{j}}, \alpha_j\in \N)$ are independent Poisson processes with density~$\exp(C_jx)dx$ and $C_1<C_2\dots<C_k$.

The 2-replica spin overlap distribution $x(q)$ can then be written (using an argument similar to that in~\cite{supp})
\begin{equation}
\label{eq:k4}
x(q)=\lambda_1 1_{[q_1,q_2)}+ \lambda_2 1_{[q_2,q_3)}+\dots +\lambda_k1_{[q_k,1)}(q)\, ,
\end{equation}
where $0\leq q_1 <\dots<q_l<\dots \leq q_{k+1} =1$, $0<\lambda_1<\dots<\lambda_k<1$, and $1_{[a,b)} = 1$ for $q\in[a,b)$ and is zero otherwise.

The map $W_\alpha\mapsto p_\alpha^{(n)}$ induces the following map on $x(q)$:
\begin{equation}
\label{eq:k5}
x(q)\mapsto x^{(n)}(q):={\lambda_1\over n} 1_{[q_1,q_2)}+ {\lambda_2\over n} 1_{[q_2,q_3)}+\dots
+{\lambda_k\over n}1_{[q_k,1)}(q)\, .
\end{equation}
From this one recovers a linear dependence on $1/n$ for ~$x^{(n)}(q)$. In particular, 
\begin{equation}
\label{eq:k6}
\int_0^1x(q) \ dq= n\int_0^1x^{(n)}(q) \ dq\, 
\end{equation}
showing that the consistency requirement imposed by~(\ref{eq:simplified}) on weights and overlaps for all $n$ is maintained for all $k$
in the Derrida-Ruelle cascade if the weights are PD-distributed.

We considered above the $\epsilon>0$ case.  The procedure for $\epsilon=0$ is identical, with the
substitutions $\Delta h_x \to \Delta J_{xy}$,
$\langle\sigma_x\rangle_\alpha\to
\langle\sigma_x\sigma_y\rangle_\alpha$, and with averages over sites $x$ 
replaced by averages over edges $\langle xy\rangle$. With these transformations, all consistency relations are unchanged except that spin
overlaps are replaced by edge overlaps. Consequently the conclusions for $\epsilon>0$ carry
over to $\epsilon=0$.

{\it Discussion.}   We have presented a method for generating consistency relations that must be satisfied in any finite dimension by any set of thermodynamic states on the spin glass metastate.
An important feature of this method, which it shares with~\cite{AD11}, is that it is not restricted to Gaussian couplings (any continuous distribution with finite mean and variance 
suffices) and holds at any fixed temperature. Moreover, it acts directly on the EA~Hamiltonian, as opposed to adding $p$-spin perturbations to it. We have presented only relations 
derived by varying the couplings or fields up to second order; however, the process can be used up to derivatives of any order. A thermodynamic identity obtained by taking
the $n^{\rm th}$ derivative of a translation-invariant function will yield relations between weights, $n$-spin (or edge) overlaps, and/or $n$-spin (or edge) correlation functions.
All such relations are trivially satisfied by simple scenarios like droplet-scaling or chaotic pairs (and also by the high-temperature paramagnetic phase), but they impose strong constraints on nontrivial mixed-state pictures.

It is natural to ask whether this method provides an alternate means of deriving Ghirlanda-Guerra identities~\cite{GG98} or stochastic stability relations~\cite{CG13,AC98,AD11}.
We discuss this briefly in the Supplementary Notes~\cite{supp}, but summarize the main ideas here.  The former contain nonlinearities in the form of products of metastate averages, so in general Ghirlanda-Guerra~identities differ from those presented here (except possibly for linear combinations of these identities that remove the nonlinear terms). On the other hand, the method described here can reproduce some stochastic stability identities, but 
again the sets of identities readily derivable by the two methods appear to be largely distinct.

In the above discussion we considered a particular function~(\ref{eq:f}). We have also checked consistency for two other translation-invariant, measurable functions,
which lead to different identities, one for each $n$: $(1/n)W_\alpha^n$ and $(1/n)q_\Gamma^n$, where again $n$ is any real number greater than one. In the first case, 
we checked the 1-step~\cite{supp} and 2-step solutions for every $n$ and in the second case the 1-step solution for $n=2$. In all cases a PD~distribution for the weights satisfied the consistency relations.

Of course, some choice of function could lead to an inconsistency between the resulting identities and Derrida-Ruelle cascades with PD distributions on the weights.
The point of this paper is that the relations studied provide evidence that the {\it only\/} potentially self-consistent nontrivial mixed-state picture is one where overlaps are generated via Derrida-Ruelle cascades and in which weights are distributed by a Poisson-Dirichlet point process. While other pictures could conceivably exist, the methods described here provide a powerful technique for testing their self-consistency: e.g., distributions on the weights in which the metastate average of $(\sum_\alpha W_\alpha^n)$ doesn't scale as $1/n$ can already be ruled out. It is likely, given the number and complexity of
these relations, that no other nontrivial mixed state pictures can exist. This may be provable, by studying an infinite hierarchy of metastate averages on products of weights raised to different powers, and will be investigated in future work.

\begin{acknowledgments}
The authors thank Pierluigi Contucci and Nicholas Read for helpful comments on the manuscript. DLS thanks the John Simon Guggenheim Foundation for a fellowship that
partially supported this research, the Aspen Center for Physics (under
NSF Grant 1066293) where part of this work was done, 
and Grace Tyrrell and Laura Stein for helpful conversations. This research was partially supported by NSF
Grant~DMS-1513441 and PSC-CUNY Research Award~68784-00~46 (LPA), and by NSF Grant~DMS-1207678 (CMN and DLS).
\end{acknowledgments}


\begin{thebibliography}{99}
\bibitem{notemany}
 This should not be confused with the presence or absence of many pure states. For example, there could in principle be a single thermodynamic state consisting of
 a continuum mixture of uncountably many pure states.
\end{thebibliography}


\end{document}


\draft

\title{Supplementary Notes for ``Thermodynamic Identities and Symmetry Breaking in Short-Range Spin Glasses''}

\author{L.-P.~Arguin}
\affiliation{Department of Mathematics, City University of New York, Baruch College and Graduate Center, New York, NY 10010}
\author{C.M.~Newman}
\affiliation{Courant Institute of Mathematical Sciences, New York, NY 10012 USA and NYU-ECNU Institute of Mathematical Sciences at
NYU Shanghai, 3663 Zhongshan Road North, Shanghai 200062, China}
\author{D.L.~Stein}
\affiliation{Department of Physics and Courant Institute of Mathematical Sciences,
New York University, New York, NY 10012 USA and NYU-ECNU Institutes of Physics and Mathematical Sciences at NYU Shanghai, 3663 Zhongshan Road North, Shanghai, 200062, China}

\pacs{}
\maketitle

\section{2-step RSB for $f(\Gamma)={1\over n}\log\Bigl(\sum_\alpha W_\alpha^n\Bigr)$}

At the 2-step level, each state $\alpha$ is represented by a pair $\alpha=(\alpha_1, \alpha_2)$, where $\alpha_1$ is the state at the first level and $\alpha_2$ the state at the second level within $\alpha_1$. The weight $W_\alpha$ is then
\begin{equation}
\label{eq:W}
W_\alpha=\frac{\eta_{\alpha_1}\eta_{\alpha_1,\alpha_2}}{\sum_{\gamma_1,\gamma_2}\eta_{\gamma_1}\eta_{\gamma_1,\gamma_2}}\, ,
\end{equation}
where $(\eta_{\alpha_1})$ is a Poisson process on $[0,\infty)$ with density $\lambda_1s^{-\lambda_1-1}ds$ and for each $\alpha_1$, $(\eta_{\alpha_1,\alpha_2},\alpha_2\in\N)$ are independent Poisson process on $[0,\infty)$ with density $\lambda_2s^{-\lambda_2-1}ds$. We must also have $\lambda_2>\lambda_1$. 
The overlap structure is then
\begin{equation}
\label{eq:q}
\bar q^{\alpha\beta}=\begin{cases}
1 &\mbox{if $\alpha=\beta$}\\
q_2&\mbox{if $\alpha_1=\beta_1$ but $\alpha_2\neq\beta_2$}\\
q_1&\mbox{if $\alpha_1\neq\beta_1$ }\, .
\end{cases}
\end{equation}
From this we find 
\begin{equation}
\label{eq:step1}
\E\Big[\sum_{\alpha,\beta}p^{(n)}_\alpha p^{(n)}_\beta \delta(q^{\alpha\beta}-1)\Big]=
\E\Big[\sum_{\alpha_1}(p^{(n)}_{\alpha_1})^2 \sum_{\alpha_2}(p^{(n)}_{\alpha_1\alpha_2})^2\Big]=1-\frac{\lambda_2}{n}\, ,
\end{equation}
\begin{equation}
\label{eq:step2}
\E\Big[\sum_{\alpha,\beta}p^{(n)}_\alpha p^{(n)}_\beta \delta(q^{\alpha\beta}-q_2)\Big]=
\E\Big[\sum_{\alpha_1}(p^{(n)}_{\alpha_1})^2 \sum_{\alpha_2\neq\beta^2}p^{(n)}_{\alpha_1\alpha_2}p^{(n)}_{\alpha_1\beta_2}\Big]=(1-\frac{\lambda_1}{\lambda_2})(\frac{\lambda_2}{n})
=\frac{\lambda_2}{n}-\frac{\lambda_1}{n}\, ,
\end{equation}
and
\begin{equation}
\label{eq:step3}
\E\Big[\sum_{\alpha,\beta}p^{(n)}_\alpha p^{(n)}_\beta \delta(q^{\alpha\beta}-q_1)\Big]=
\E\Big[\sum_{\alpha_1\neq \beta_1}p^{(n)}_{\alpha_1}p^{(n)}_{\beta_1}\Big]=\frac{\lambda_1}{n}\, .
\end{equation}

Therefore, the metastate average on the RHS of~(16) in the text is
\begin{equation}
\label{eq:qav}
\E\Big[\sum_{\alpha,\beta}p^{(n)}_\alpha p_\beta^{(n)} \bar{q}^{\alpha\beta}\Big]
=q_1\frac{\lambda_1}{n} + q_2\left(\frac{\lambda_2}{n}-\frac{\lambda_1}{n}\right)+ 1-\frac{\lambda_2}{n}
\end{equation}
On the other hand,
\begin{equation}
\label{eq:other hand}
\E[\bar q_\Gamma]=q_1\lambda_1 + q_2\left(\lambda_2-\lambda_1\right)+ (1-\lambda_2)\, .
\end{equation}
Therefore after simplification, Eq.~(16) reduces to
\begin{equation}
\label{eq:check}
1-q_1\lambda_1 - q_2 (\lambda_2-\lambda_1) - (1-\lambda_2)=n\Big(1-q_1\frac{\lambda_1}{n} - q_2(\frac{\lambda_2}{n}-\frac{\lambda_1}{n}) - (1-\frac{\lambda_2}{n}))
\end{equation}
and the set of consistency relations is satisfied for all $n$.

\vfill\eject

\section{1-step RSB for $f(\Gamma)={1\over n}\sum_\alpha W_\alpha^n$}

We consider the translation-invariant function ${\cal F}=\frac{1}{n}\sum_\alpha W_\alpha^n$, which results in the identities
\begin{equation}
\label{eq:intderiv}
E_x\Bigl[{\partial_{xx} \int d\lambda(\Gamma)\ {\cal F}(\Gamma)}\Bigr]=\int d\lambda(\Gamma)\ \sum_\alpha W_\alpha^n\Bigl[(n-1)q_{EA} -2nq^\alpha +(n+1)q_\Gamma\Bigr]\, ,
\end{equation}
where $E_x$ is an average over all sites, $q_\Gamma$ is defined as before and $q^\alpha=\sum_\beta W_\beta q^{\alpha\beta}$ is the average overlap of pure state
$\alpha$ with all the other pure states in $\Gamma$.

For 1-step RSB, $q^{\alpha\alpha}=q_{EA}$ and $q^{\alpha\beta}=q_1$ for $\alpha\ne\beta$, with $q_1<q_{EA}$.
In this case the following relations hold:
\begin{equation}
\label{eq:qgamma}
q_\Gamma=\sum_\alpha W_\alpha^2 q_{EA} + (1-\sum_\alpha W_\alpha^2)q_1 =  q_1 + \sum_\alpha W_\alpha^2(q_{EA}-q_1)
\end{equation}
and
\begin{equation}
\label{eq:qalpha}
q^\alpha = W_\alpha q_{EA} + (1-W_\alpha)q_1 = q_1 +W_\alpha(q_{EA}-q_1)\, .
\end{equation}

Inserting~(\ref{eq:qgamma}) and~(\ref{eq:qalpha}) into~(\ref{eq:intderiv}), we find
\begin{equation}
\label{eq:condition}
E_x\Bigl[{\partial_{xx} \int d\kappa(\Gamma)\ {\cal F}(\Gamma)}\Bigr]=(q_{EA}-q_1)\E\Bigl[(n-1)\sum_\alpha W_\alpha^n -2n \sum_\alpha W_\alpha^{n+1} +(n+1)\sum_\alpha W_\alpha^n\sum_\alpha W_\alpha^2\Bigr]\, ,
\end{equation}

If the $W_\alpha$'s are PD($\lambda$), then using Eq.~(15) from the main text we have
\begin{equation}
\label{eq:exp1}
\E\Bigl[\sum_\alpha W_\alpha^n\Bigr]=\frac{\Gamma(n-\lambda)}{\Gamma(1-\lambda)\Gamma(n)}\, .
\end{equation}
To evaluate the third term on the RHS we use the relation presented in~\cite{BS98}
\begin{equation}
\label{eq:product}
\E\left[\sum_{\alpha_1,\ldots,\alpha_m}W_{\alpha_1}^{n_1}\ldots W_{\alpha_m}^{n_m}\right]=\frac{(m-1)!}{(n-1)!}\lambda^{m-1}\prod_{k=1}^m(n_k-1-\lambda)\ldots(1-\lambda)\, ,
\end{equation}
where $n=\sum_{k=1}^m n_k$ and the pure states $(\alpha_1,\ldots,\alpha_m)$ are all distinct.

We then find
\begin{equation}
\label{eq:exp2}
\E\Bigl[\sum_\alpha W_\alpha^n\sum_\alpha W_\alpha^2\Bigr]=\frac{1}{\Gamma(n+2)\Gamma(1-\lambda)}\Bigl[\Gamma(n+2-\lambda)+\lambda(1-\lambda)\Gamma(n-\lambda)\Bigr]\, .
\end{equation}
Substituting~(\ref{eq:exp1}) and~(\ref{eq:exp2}) into~(\ref{eq:condition}), one finds that the RHS is zero for every $n$.

\section{Relation to Ghirlanda-Guerra and stochastic stability relations}

A natural question concerns the relationship of the method proposed in this
paper to those leading to the well-known Ghirlanda-Guerra~\cite{GG98} and
stochastic stability~\cite{AC98,CG13} identities.  This is a subject of
current investigation, and here we present only some preliminary remarks.

All three methods follow from fundamental invariance principles of the
statistical mechanics of systems with quenched disorder. The
Ghirlanda-Guerra identities are a consequence of the vanishing of
fluctuations of the energy per particle in the thermodynamic limit; the
stochastic stability identities follow from the invariance of the quenched
expectation of a bounded function on multiple copies of the spins when the
Hamiltonian is perturbed by an independent Gaussian process; and identities
derived using the method presented here follow from the constancy of a
translation-invariant function on the disorder with respect to finite
changes in the couplings and/or fields. The Ghirlanda-Guerra identities
include as a subset those derivable using stochastic stability (though in
practice some identities may be easier to derive using stochastic
stability). At least some stochastic stability identities can be recovered using the
approach of this paper following the method presented in~\cite{AD11}.
Whether some or all of the identities derived by the method presented in
this paper are derivable using the Ghirlanda-Guerra method, or vice-versa,
is unknown at this time but presents an interesting subject for further
investigation.

We first discuss some important differences. Both Ghirlanda-Guerra and
stochastic stability identities are usually presented as averages over the
temperature, although this can sometimes be avoided by adding $p$-spin
perturbations to the Hamiltonian. In the method presented here, relations
holding at a single fixed temperature are obtained directly from the
unperturbed Hamiltonian~\cite{note}. A second difference is that the method
used here works for any continuous coupling distribution with finite
variance, while in the usual formulation the Ghirlanda-Guerra and
stochastic stability identities are restricted to Gaussian
couplings. (However, an extension of these methods to other coupling
distributions was presented in~\cite{AD11}.) A third difference is that the
method presented here relates weights to both overlaps and correlation
functions, while the older methods, at least in their most natural
formulation, give rise to relations among functions on overlaps and
weights.

We have found that some identities are derivable using any of these methods. Consider, for example, the identity
\begin{equation}
\label{eq:identity}
\mathbb{E}\big[{\rm Var}_\Gamma(q^{\alpha\beta})\big]=4\mathbb{E}\big[{\rm Var}_\Gamma(q^{\alpha})\big]\, ,
\end{equation}
where $q^\alpha=\sum_\beta W_\beta q^{\alpha\beta}$ is the average overlap in the mixed state $\Gamma$ of the pure state $\alpha$ with all other pure states in $\Gamma$.

This is an interesting identity. It can derived using the method in the main text by taking a second derivative with respect to $h_x$ of the translation-invariant function $q_\Gamma$, the average spin overlap in $\Gamma$ (Eq.~(4) in the main text), and then averaging over all sites. But it is also equivalent to the identities~(5.89) and~(5.90) in~\cite{CG13}, derived using stochastic stability. So as noted above at least some identities can be derived using any of these methods. To what extent this holds for a more general class of identities is a topic for further investigation.